\documentclass{stacs_proc}
\usepackage{euscript}
\usepackage{graphicx}

\DeclareMathSymbol{\upset}{\mathopen}{symbols}{"22}
\newcommand{\supp}{\mathrm{supp}}
\newcommand{\scr}{\EuScript}
\renewcommand{\colon}{:}

\newenvironment{pf}{\proof}{\qed}

\begin{document}
\title[Trimmed Moebius Inversion and Graphs of Bounded
Degree]{Trimmed Moebius Inversion and \\
Graphs of Bounded Degree}

\thanks{This research was supported in part by the Academy of
Finland, Grants 117499 (P.K.) and 109101 (M.K.).}

\author[lu]{A. Bj\"orklund}{Andreas Bj\"orklund}
\author[lu]{T. Husfeldt}{Thore Husfeldt}

\address[lu]{Lund University, Department of Computer Science, P.O.Box 118, SE-22100 Lund, Sweden}
\email{andreas.bjorklund@logipard.com, thore.husfeldt@cs.lu.se}  

\author[he]{P. Kaski}{Petteri Kaski}

\author[he]{M. Koivisto}{Mikko Koivisto}

\address[he]{Helsinki Institute for Information Technology HIIT,
  University of Helsinki, Department of Computer Science, P.O.Box 68,
  FI-00014 University of Helsinki, Finland}
\email{\{petteri.kaski,mikko.koivisto\}@cs.helsinki.fi}  

\renewcommand{\url}[1]{#1}

\begin{abstract}
  We study ways to expedite Yates's algorithm for computing the zeta
  and Moebius transforms of a function defined on the subset
  lattice. We develop a trimmed variant of Moebius inversion that
  proceeds point by point, finishing the calculation at a subset 
  before considering its supersets.
  For an $n$-element universe $U$ and a family $\scr F$ of its subsets,
  trimmed Moebius inversion allows us to compute the number of packings, 
  coverings, and partitions of $U$ with $k$ sets from $\scr F$ in time 
  within a polynomial factor (in $n$) of the number of supersets of 
  the members of $\scr F$.

  Relying on an intersection theorem of Chung \emph{et al.} (1986) to bound the sizes of set families, we apply 
  these ideas to well-studied combinatorial optimisation problems 
  on graphs of maximum degree $\Delta$. In particular, 
  we show how to compute the Domatic Number in time within 
  a polynomial factor of $(2^{\Delta+1}-2)^{n/(\Delta+1)}$ and 
  the Chromatic Number in time within a polynomial factor of 
  $(2^{\Delta+1}-\Delta-1)^{n/(\Delta+1)}$. For any constant $\Delta$,
  these bounds are $O\bigl((2-\epsilon)^n\bigr)$ for $\epsilon>0$
  independent of the number of vertices $n$.
\end{abstract}

\maketitle

\stacsheading{2008}{85-96}{Bordeaux}
\firstpageno{85}

\vskip-0.3cm
\section{Introduction}

Yates's algorithm from 1937 is a kind of fast Fourier transform that
computes for a function $f\colon \{0,1\}^n\rightarrow\mathbf R$ and another
function $\upsilon\colon \{0,1\}\times\{0,1\}\rightarrow\mathbf R$ the values
\begin{equation}
  \label{eq: yates sum intro} \widehat f(x_1,\ldots, x_n)\ =\ 
  \sum_{y_1,\ldots,y_n \in\{0,1\}}\ 
  \upsilon(x_1,y_1)\cdots
  \upsilon(x_n,y_n)
  f(y_1,\ldots,y_n)\,.
\end{equation}
simultaneously for all $X=(x_1,\ldots,x_n)\in\{0,1\}^n$ using 
only $O(2^nn)$ operations, instead of the obvious $O(4^nn)$. 
The algorithm is textbook material in many sciences. Yet, though it appears in Knuth \cite[\S 3.2]{Knuth2}, 
it has received little attention in combinatorial optimisation. 

Recently, the authors \cite{BHKK07, BHK08} used Yates's
algorithm in combination with Moebius inversion to give algorithms for
a number of canonical combinatorial optimisation problems such as
Chromatic Number and Domatic Number in $n$-vertex graphs, and
$n$-terminal Minimum Steiner Tree, in running times within a
polynomial factor of $2^n$.

From the way it is normally stated, Yates's algorithm seems to face an
inherent $2^n$ lower bound, up to a polynomial factor, and it
also seems to be oblivious to the structural properties of the
transform it computes.

The motivation of the present investigation is to expedite the running
time of Yates's algorithm for certain structures so as to get running
times with a dominating factor of the form $(2-\epsilon)^n$. 
From the perspective of running times alone, our improvements are 
modest at best, but apart from providing evidence that the 
aesthetically appealing $2^n$ bound from \cite{BHK08} 
can be beaten, the combinatorial framework we present seems to be 
new and may present a fruitful direction for exact exponential 
time algorithms.

\subsection{Results}
In a graph $G=(V,E)$, a set $D\subseteq V$ of vertices is
\emph{dominating} if every vertex not in $D$ has at least one
neighbour in $D$. The \emph{domatic number} of $G$ is the largest $k$
for which $V$ can be partitioned in to $k$ dominating sets.  We show
how to compute the domatic number of an $n$-vertex graph with maximum
degree $\Delta$ in time 
\[
O^*\bigl((2^{\Delta+1}-2)^{n/(\Delta+1)}\bigr)\,;
\] 
the $O^*$ notation suppresses factors that are polynomial in $n$.  
For constant $\Delta$, this bound is always better than $2^n$, 
though not by much:

\medskip
\centerline{\small
\begin{tabular}{c|ccccccc}
  $\Delta$  & 3 & 4 & 5 & 6 & 7 & 8 & $\cdots$ \\\hline
\rule{0ex}{1.5em}$(2^{\Delta+1}-2)^{1/(\Delta+1)}$ & 
$1.9344$ & $1.9744$ & $1.9895$ & $1.9956$ & $1.9981$ & $1.9992$ & $\cdots$
\end{tabular}}
\medskip

The \emph{chromatic number} of a graph is the minimum $k$ for which 
the vertex set can be covered with $k$ independent sets; 
a set $I\subseteq V$ is \emph{independent} if no two vertices in $I$ 
are neighbours.
We show how to compute the chromatic number of an $n$-vertex graph 
with maximum degree $\Delta$ in
time 
\[
O^*\bigl((2^{\Delta+1}-\Delta-1)^{n/(\Delta+1)}\bigr)\,. 
\]
This is slightly faster than for Domatic Number:

\medskip
\centerline{\small
\begin{tabular}{c|ccccccc}
  $\Delta$  & 3 & 4 & 5 & 6 & 7 & 8 & $\cdots$\\\hline
 \rule{0ex}{1.5em} $(2^{\Delta+1}-\Delta-1)^{1/(\Delta+1)}$ &  $1.8613$ &
  $1.9332$ & 
  $1.9675$ & 
  $1.9840$ & 
  $1.9921$ & 
  $1.9961$ & 
$\cdots$
\end{tabular}}
\medskip

One notes that even for moderate $\Delta$, the improvement over $2^n$
is minute. Moreover, the colouring results for $\Delta\leq 5$ are not
even the best known: by Brooks's Theorem \cite{Brooks41}, 
the chromatic number of a connected graph is bounded by its
maximum degree unless the graph is complete or an odd cycle, 
both of which are easily recognised. It remains to decide if the 
chromatic number is 3, 4, or 5, and with algorithms from the literature, 
$3$- and $4$-colourability can be decided in time $O(1.3289^n)$ 
\cite{BeiEpp05} and $O(1.7504^n)$ \cite{Bys04}, respectively. 
However, this approach does stop at $\Delta=5$, since we know 
no $o(2^n)$ algorithm for $5$-colourability. Other approaches for 
colouring low-degree graphs are known via pathwidth:
given a path decomposition of width $w$ the $k$-colourability can be
decided in time $k^w n^{O(1)}$ \cite{FGSS06};  
for $6$-regular graphs one can find a decomposition with 
$w \leq n(23+\epsilon)/45$ for any $\epsilon>0$ and
sufficiently  large $n$ \cite{FGSS06}, and for graphs with
$m$ edges one can find $w \leq m/5.769 + O(\log n)$ \cite{KMRR05}. 
However, even these pathwidth based bounds fall short when 
$k \geq 5$---we are not aware of any previous $o(2^n)$ algorithm.
 
For the general case, it took 30 years and many papers to improve the
constant in the bound for Chromatic Number from $2.4423$~\cite{Law76}
via $2.4151$~\cite{Epp03}, $2.4023$~\cite{Bys04},
$2.3236$~\cite{BH08}, to~$2$~\cite{BHK08}, and a
similar (if less glacial) story can be told for the Domatic
Number. None of these approaches was sensitive to the density of the
graph. Moreover, what interests us here is not so much the size of the
constant, but the fact that it is less than $2$, dispelling the tempting 
hypothesis that $2^n$ should be a `difficult to beat' bound for computing 
the Chromatic Number for sparse graphs. 
In \S\ref{sec:
 con} we present some tailor-made variants for which the 
running time improvement from applying the ideas of the present paper 
are more striking.

Chromatic Number and Domatic Number are special cases of set
partition problems, where the objective is to partition an $n$-element 
set $U$ (here, the vertices of a graph) with members of a given family
$\scr F$ of its subsets (here, the independent or dominating sets of 
the graph). In full generality, we show how to compute the covering,
packing, and partition numbers of $(U,\scr F)$ in time within a
polynomial factor of 
\begin{equation}
\label{eq: running time}
|\{ T\subseteq U \colon 
    \text{there exists an $S\in \scr F$ such that $S\subseteq T$}\}|\,, 
\end{equation} 
the number of supersets of the members of $\scr F$. In the worst case,
this bound is not better than $2^n$, and the combinatorial
challenge in applying the present ideas is to find good bounds
on the above expression.

\subsection{Techniques}
The main technical contribution in this paper, sketched in
Figure~1, is that Yates's algorithm can,
for certain natural choices of 
$\upsilon\colon \{0,1\}\times\{0,1\}\rightarrow\mathbf R$, be trimmed by
considering in a bottom-up fashion only those $X\in\{0,1\}^n$ that we
are actually interested in, for example those $X$ for which $f(X)\neq
0$ and their supersets. (We will understand $X$ as a subset of
$\{1,\ldots,n\}$ whenever this is convenient.) Among the
transforms that are amenable to trimming are the zeta and Moebius 
transforms on the subset lattice. 

\begin{figure}
\parbox{4cm}{\begin{tabular}{c}\includegraphics{trim.1}
\includegraphics{trim.2}
\includegraphics{trim.3}\end{tabular}}
\parbox{11cm}{\small Figure 1: Trimmed evaluation. Originally, Yates's
  algorithm considers the entire subset lattice (left). We trim the
  evalation from below by considering only the supersets of
  `interesting' points (middle), and from above by abandoning computation
  when we reach certain points (right).}
\end{figure}

We use the trimmed algorithms for zeta and Moebius transforms
to expedite Moebius inversion, a generalisation of the principle 
of inclusion--exclusion, which allows us to compute the cover, packing, 
and partition numbers. The fact that these numbers can be 
computed via Moebius inversion was already used in
\cite{BH08,BHKK07,BHK08}, and those parts of the
present paper contain little that is new, except for a somewhat more
explicit and streamlined presentation in the framework of partial
order theory.

The fact that we can evaluate both the zeta and Moebius transforms 
pointwise in such a way that we are done with $X$ before we proceed to 
$Y$ for every $Y\supset X$ also enables us to further trim computations 
from what is outlined above. For instance, if we seek a minimum set 
partition of sets from a family $\scr F$ of subsets of $U$, then it 
suffices to find the minimum partition of all $X$ such that $U\setminus X=S$ 
for some $S\in \scr F$. In particular, we need not consider how many 
sets it takes to partition $X$ for $X$'s large enough for $U\setminus X$ not 
to contain any set from $\scr F$.

The main combinatorial contribution in this paper is that if
$\scr F$ is the family of maximal independent sets, or the family of
dominating sets in a graph, then we show how to bound 
\eqref{eq: running time} in terms of the maximum degree $\Delta$ using 
an intersection theorem of Chung \emph{et al.} \cite{CFGS86} 
that goes back to Shearer's Entropy Lemma. For this we merely need to 
observe that the intersection of $\scr F$ and the closed neighbourhoods 
of the input graph excludes certain configurations.

In summary, via \eqref{eq: running time} the task of bounding the
running time for (say) Domatic Number reduces to a combinatorial
statement about the intersections of certain families of sets.

\subsubsection*{Notation}
Yates's algorithm operates on the lattice of subsets of an
$n$-element universe $U$, and we find it convenient to work with 
notation established in partial order theory.

For a family $\scr F$ of subsets of $U$, let 
$\min\scr F$ (respectively, $\max\scr F$) denote the family of 
minimal (respectively, maximal) elements of $\scr F$ with respect to 
subset inclusion. The \emph{upper closure} 
(sometimes called \emph{up-set} or \emph{filter}) of $\scr F$ is 
defined as
\[
\upset\scr F = \{\, T\subseteq U\colon  
                     \text{there exists an $S\in \scr F$ such that $S\subseteq T$}\,\}\,.
\]
For a function $f$ defined on subsets of $U$, the \emph{support} of $f$
is defined as
\[
\supp(f)=\{\,X\subseteq U\colon f(X)\neq 0\}\,.
\]

For a graph $G$, 
we let $\scr D$ denote the family of \emph{dominating sets} of $G$
and $\scr I$ the family of \emph{independent sets} of $G$.
Also, for a subset $W\subseteq V$ of vertices, 
we let $G[W]$ denote the subgraph induced by $W$.
For a proposition $P$, we use Iverson's bracket notation
$[P]$ to mean $1$ if $P$ is true and $0$ otherwise.

\section{Trimmed Moebius Inversion}

For a family $\scr F$ of sets from $\{0,1\}^n$ and a set $X\in
\{0,1\}^n$ we will consider $k$-tuples $(S_1,\ldots,S_k)$ with
$S_i\in\scr F$ and $S_i\subseteq X$. Such a tuple is \emph{disjoint}
if $S_{i_1}\cap S_{i_2}=\emptyset$ for all $1\leq i_1<i_2\leq k$, and 
\emph{covering} if $S_1\cup\cdots \cup S_k= X$. From these concepts we
define for fixed $k$
\begin{enumerate}
\item the \emph{cover number} $c(X)$, viz.\ the number of covering
  tuples,
\item the \emph{packing number} $p(X)$, viz.\ the number of disjoint
  tuples,
\item the \emph{partition number} or \emph{disjoint cover number}
  $d(X)$, viz.\ the number of tuples that are both disjoint and covering.
\end{enumerate}

In this section we show how to compute these numbers in time
$|\upset\scr F|n^{O(1)}$, rather than $2^nn^{O(1)}$ as in
\cite{BHKK07, BHK08}.  The algorithms are concise but
somewhat involved, and we choose to present them here starting with an
explanation of Yates's algorithm. Thus, the first two subsections are
primarily expository and aim to establish the new ingredients in our
algorithms.

At the heart of our algorithms lie two transforms of functions
$f\colon \{0,1\}^n\rightarrow \mathbf R$ on the subset lattice. The
\emph{zeta} transform $f\zeta$ is defined for all $X\in\{0,1\}^n$ by
\begin{equation}\label{eq: zeta transform}
   (f\zeta)(X)\,=\sum_{Y\subseteq X} f(Y)\,.
\end{equation}
(The notation $f\zeta$ can be read either as a formal operator or as
a product of the $2^n$-dimensional vector $f$ and the matrix
$\zeta$ with entries $\zeta_{YX}=[Y\subseteq X]$.)
The \emph{Moebius} transform $f\mu$ is defined  
for all $X\in\{0,1\}^n$ by
\begin{equation}\label{eq: moebius transform}
  (f\mu)(X)\,=\sum_{Y\subseteq X}(-1)^{|X\setminus Y|} f(Y)\,.
\end{equation} 
These transforms are each other's inverse in the sense that
$f=f\zeta\mu=f\mu\zeta$, a fundamental combinatorial principle called
\emph{Moebius inversion}. We can (just barely) draw an example in four
dimensions for a function $f$ given by  $f(\{4\})=f(\{1,2,4\})=1$,
$f(\{1,3\})=2$ and $f(X)=0$ otherwise: 

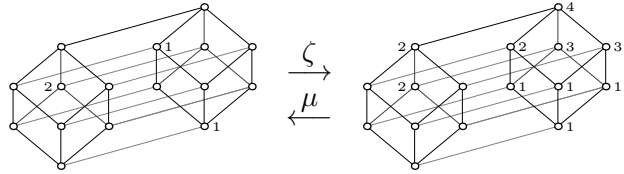
\[\vcenter{\hbox{\includegraphics{lattices.1}}}
\quad \begin{matrix}
\stackrel{\textstyle\zeta}{\longrightarrow}\\
\stackrel{\textstyle\mu}{\longleftarrow}
\end{matrix}
\quad
\vcenter{\hbox{\includegraphics{lattices.2}}}\]
Another example that we will use later is the connection between
the packing number and the disjoint cover number,
\begin{equation}
\label{eq: p=dzeta} 
p=d\zeta\,, 
\end{equation}
which is easy to verify: By definition, 
\[
(d\zeta)(X)\,=\sum_{Y\subseteq X} d(Y)\,.
\] 
Every disjoint $k$-tuple
$(S_1,\ldots,S_k)$ with $S_1\cup\cdots\cup S_k\subseteq X$ appears
once on the right hand side, namely for $Y=S_1\cup\cdots\cup S_k$, so
this expression equals the packing number $p(X)$.

\subsection{Yates's algorithm} 
Yates's algorithm \cite{Yates1937} expects the transform in the form
of a function $\upsilon\colon \{0,1\}\times\{0,1\}\rightarrow\mathbf R$ and
computes the transformed values
\begin{equation}
  \label{eq: yates sum} 
  \widehat f(X)\ = \sum_{Y\in\{0,1\}^n} 
                     \upsilon(x_1,y_1)\cdots\upsilon(x_n,y_n) f(Y)\,.
\end{equation}
simultaneously for all $X\in\{0,1\}^n$. Here, we let $(x_1,\ldots,x_n)$ and $(y_1,\ldots,y_n)$ denote the binary representations (or,
`incidence vectors') of $X$ and $Y$, so $x_j=[j\in X]$ and $y_j=[j\in Y]$. 
To obtain \eqref{eq: zeta transform} set $\upsilon(x,y)=[y\leq x]$ and to
obtain \eqref{eq: moebius transform} set $\upsilon(x,y)=[y\leq x](-1)^{x-y}$. 
  
The direct evaluation of \eqref{eq: yates sum} would take $2^n$
evaluations of $f$ for each $X$, for a total of $O(2^n2^nn)=O(4^nn)$
operations. The zeta and Moebius transforms depend only on $Y\subseteq X$, 
so they would require only $\sum_{X}2^{|X|}=\sum_{0\leq i\leq n}\binom{n}{i} 2^i=3^n$ evaluations.  Yates's algorithm
is faster still and computes the general form in $O(2^nn)$ operations:

\medskip\noindent{\small {\bf Algorithm~Y.} (\emph{Yates's algorithm.}) Computes
$\widehat f(X)$ defined in \eqref{eq: yates sum} for all $X\in\{0,1\}^n$ 
given $f(Y)$ for all $Y\in \{0,1\}^n$ and
$\upsilon(x,y)$ for all $x,y\in\{0,1\}$.
\begin{itemize}
\item[{\bf Y1:}] For each $X\in\{0,1\}^n$, set $g_0(X)=\!f(X)$.
\item[{\bf Y2:}] For each $j=1,\ldots,n$ and $X\in\{0,1\}^n$, set
  \[
     g_j(X)= 
       \upsilon([j\in X],0) g_{j-1}(X\setminus \{j\})+
       \upsilon([j\in X],1) g_{j-1}(X\cup \{j\})\,.
  \]
\item[{\bf Y3:}] Output $g_n$.
\end{itemize}}

The intuition is to compute $\widehat f(X)$ `coordinate-wise' by fixing fewer
and fewer bits of $X$ in the sense that, for $j=1,\ldots, n$,
\begin{equation}
\label{eq: y induction}
g_j(X)\ = \sum_{y_1,\ldots, y_j\in\{0,1\}}\,\upsilon(x_1,y_1)\cdots
\upsilon(x_j,y_j) f(y_1,\ldots,y_j,x_{j+1},\ldots,x_n)\,.
\end{equation}
Indeed, the correctness proof is a straightforward verification 
(by induction) of the above expression.

\subsection{Trimmed pointwise evaluation}
To set the stage for our present contributions, observe that both the 
zeta and Moebius transforms `grow upwards' in the subset lattice in 
the sense that $\supp(f\zeta),\supp(f\mu)\subseteq \upset \supp(f)$. 
Thus, in evaluating the two transforms, one ought to be able 
to trim off redundant parts of the lattice and work only with lattice 
points in $\upset\supp(f)$. 

We would naturally like trimmed evaluation to occur in $O(|\upset\supp(f)|n)$ 
operations, in the spirit of Algorithm~Y. However, to obtain the values 
at $X$ in Step Y2 of Algorithm~Y, at first sight it appears that we 
must both `look up' (at $X\cup\{j\}$) and 'look down' (at $X\setminus\{j\}$). 
Fortunately, it suffices to only `look down'.
Indeed, for the zeta transform, setting $\upsilon(x,y)=[y\leq x]$ and 
simplifying Step Y2 yields 
\begin{equation}
\label{eq: y zeta recursion}
     g_j(X)= 
       [j\in X] g_{j-1}(X\setminus \{j\})+g_{j-1}(X)\,.
\end{equation}
For the Moebius transform, setting $\upsilon(x,y)=[y\leq x](-1)^{x-y}$ 
and simplifying yields
\begin{equation}
\label{eq: y moebius recursion}
     g_j(X)= -[j\in X] g_{j-1}(X\setminus \{j\})+g_{j-1}(X)\,.
\end{equation}
Furthermore, it is not necessary to look `too far' down:
for both transforms it is immediate from \eqref{eq: y induction} that 
\begin{equation}
\label{eq: y support trimming}
\text{$g_j(X)=0$ holds for all $X\notin\upset\supp(f)$ and $j=0,\ldots,n$}\,.
\end{equation}
In what follows we tacitly employ (\ref{eq: y support trimming}) to
limit the scope of (\ref{eq: y zeta recursion}) and
(\ref{eq: y moebius recursion}) to $\upset\supp(f)$.

The next observation is that the lattice points in $\upset\supp(f)$ 
can be evaluated in order of their rank, using sets $\scr L(r)$ 
containing the points of rank $r$. Initially, the sets $\scr L(r)$ 
contain only $\supp(f)$, but we add elements from $\upset\supp(f)$ 
as we go along. These observations result in the following algorithm 
for evaluating the zeta transform; 
the algorithm for evaluating the Moebius transform is obtained by 
replacing \eqref{eq: y zeta recursion} in Step Z3 with 
\eqref{eq: y moebius recursion}.

\medskip\noindent{\small{\bf Algorithm~Z.} (\emph{Trimmed pointwise fast zeta
  transform.}) Computes the nonzero part of $f\zeta$ given 
               the nonzero part of $f$. 
The algorithm maintains $n+1$ families $\scr L(0),\ldots,\scr L(n)$ 
of subsets $X\in\{0,1\}^n$; $\scr L(r)$ contains only sets of size $r$. 
We compute auxiliary values $g_j(X)$ for all $1\leq j\leq n$ and
$X\in\upset\supp(f)$; it holds that $g_n(X)=(f\zeta)(X)$.
\begin{itemize}
  \item[{\bf Z1:}] For each $X\in\supp(f)$, insert $X$ into $\scr L(|X|)$. 
                   Set the current rank $r=0$.
  \item[{\bf Z2:}] Select any $X\in \scr L(r)$ and remove it from $\scr L(r)$. 
  \item[{\bf Z3:}] Set $g_0(X)=f(X)$. For each $j=1,\ldots, n$, set 
    \[ 
     g_j(X)= 
       [j\in X] g_{j-1}(X\setminus \{j\})+g_{j-1}(X)\,.
    \]
    [At this point $g_n(X)=(f\zeta)(X)$.]
  \item[{\bf Z4:}] If $g_n(X)\neq 0$, then output $X$ and $g_n(X)$.
  \item[{\bf Z5:}] For each $j\notin X$, insert $X\cup \{j\}$ into $\scr L(r+1)$.
  \item[{\bf Z6:}] If $\scr L(r)$ is empty then increment $r\leq n$ until $\scr L(r)$
    is nonempty; terminate if $r=n$ and $\scr L(n)$ is empty. 
  \item[{\bf Z7:}] Go to Z2.
\end{itemize}}

Observe that the evaluation at $X$ is complete once Step~Z3 terminates,
which enables further trimming of the lattice `from above' in case the 
values at lattice points with higher rank are not required.

By symmetry, the present ideas work just as well for transforms that 
`grow downwards', in which case one needs to `look up'.
However, they do not work for transforms that grow in both directions, 
such as the Walsh--Hadamard transform.

In the applications that now follow, $f$ will always be the indicator
function of a family $\scr F$. In this case having $\supp(f)$ quickly 
available translates to $\scr F$ being efficiently listable;
for example, with polynomial delay.

\subsection{Covers}

The easiest application of the trimmed Moebius inversion computes 
for each $X\in\upset\scr F$ the cover number $c(X)$.
This is a particularly straightforward function of the zeta transform of the
indicator function $f$: simply raise each element of $f\zeta$ to the $k$th
power and transform the result back using $\mu$.  To see this, observe
that both sides of the equation
\begin{equation}\label{eq:
    moebius for cover} (c\zeta)(Y)=
  \bigl((f\zeta)(Y)\bigr)^k
\end{equation}
count the number of ways to choose $k$-tuples $(S_1,\ldots, S_k)$ with
$S_i\subseteq Y$ and $S_i\in\scr F$.  By Moebius inversion, we can 
recover $c$ by applying $\mu$ to both sides of \eqref{eq: moebius for cover}.

\medskip\noindent{\small {\bf Algorithm~C.} (\emph{Cover number.}) Computes
$c(X)$ for all $X\in\upset\scr F$ given $\scr F$. 
The sets $\scr L(r)$ and auxiliary
values $g_j(X)$ are as in Algorithm~Z; also required are auxiliary
values $h_j(X)$ for Moebius transform.
\begin{itemize}
\item[{\bf C1:}] For each $X\in\scr F$, insert $X$ into $\scr L(|X|)$. Set the current rank
  $r=0$.
\item[{\bf C2:}] Select any $X\in \scr L(r)$ and remove it from $\scr L(r)$.
\item[{\bf C3:}] [Zeta transform.]  Set $g_0(X)=[X\in\scr F]$. For each $j=1,\ldots, n$, set 
    \[ 
       g_j(X) =    [j\in X]      g_{j-1} (X\setminus\{j\}) + 
                                 g_{j-1} (X)\,.
    \]
  [At this point it holds that $g_n(X)=(f\zeta)(X)$.]
\item[{\bf C4:}] [Evaluate zeta transform of $c(X)$.] Set $h_0(X)=g_n(X)^k$.
\item[{\bf C5:}] [Moebius transform.] For each $j=1,\ldots, n$, set 
    \[ 
       h_j(X)= -[j\in X] h_{j-1} (X\setminus\{j\}) + h_{j-1} (X)\,.
    \]
\item[{\bf C6:}] Output $X$ and $h_n(X)$.
\item[{\bf C7:}] For each $j\notin X$, insert $X\cup \{j\}$ into $\scr L(r+1)$.
\item[{\bf C8:}] If $\scr L(r)$ is empty, then increment $r\leq n$ until $\scr L(r)$
  is nonempty; terminate if $r=n$ and $\scr L(n)$ is empty. 
 \item[{\bf C9:}] Go to C2.
\end{itemize}}

\subsection{Partitions}
What makes the partition problem slightly less transparent is the fact
that we need to use dynamic programming to assemble partitions from
sets with different ranks. To this end, we need to compute for each
rank $s$ the `ranked zeta transform'
\[ 
(f\zeta^{(s)})(X)\mspace{5mu}=\mspace{-10mu}\sum_{%
Y\subseteq X, |Y|=s} f(Y)\,.
\] 

For rank $s$, consider the number $d^{(s)}(Y)$ of tuples
$(S_1,\ldots,S_k)$ with $S_i\in\scr F$, $S_i\subseteq Y$,
$S_1\cup\cdots\cup S_k=Y$ and $|S_1|+\cdots+|S_k|=s$. Then
$d(Y)=d^{(|Y|)}(Y)$. Furthermore, the zeta-transform
$(d^{(s)}\zeta)(X)$ counts the number of ways to choose
$(S_1,\ldots,S_k)$ with $S_i\subseteq X$, $S_i\in\scr F$, and
$|S_1|+\cdots+|S_k|= s$.  Another way to count the exact same quantity
is
\begin{equation}\label{eq: partition transform}
  q(k,s,X)\mspace{7mu}=\sum_{s_1+\cdots+ s_k=s}\mspace{5mu} \prod_{i=1}^k (f\zeta^{(s_i)})(X)\,.
\end{equation}
Thus we can recover $d^{(s)}(Y)$ from $q(k,s,X)$ by Moebius inversion.

As it stands, \eqref{eq: partition transform} is time-consuming to
evaluate even given all the ranked zeta transforms, but we can compute it
efficiently using dynamic programming based on the recurrence
\[ 
  q(k,s,X)=\begin{cases}
  \sum_{t=0}^s q(k-1,s-t,X)  (f\zeta^{(t)})(X),& \text{if $k>1$}\,,\\
  (f\zeta^{(s)})(X), & \text{if $k=1$}\,.\end{cases}
\]
This happens in Step~D4.

\medskip\noindent{\small {\bf Algorithm~D.} (\emph{Disjoint cover number.})
Computes $d(X)$ for all $X\in\upset\scr F$ given $\scr F$. 
The sets $\scr L(r)$ are as
in Algorithm~Z; we also need auxiliary values $g_j^{(s)}(X)$ 
and $h_j^{(s)}(X)$ for all $X\in\upset\scr F$, $1\leq j\leq n$, 
and $0\leq s\leq n$; it holds that 
$g_n^{(s)}(X)=(f\zeta^{(s)})(X)$ and $h_n^{(s)}(X)=d^{(s)}(X)$.
\begin{itemize}
\item[{\bf D1:}] For each $X\in\scr F$, insert $X$ into $\scr L(|X|)$. Set the
  current rank $r=0$.
\item[{\bf D2:}] Select any $X\in \scr L(r)$ and remove it from $\scr L(r)$.
\item[{\bf D3:}] [Ranked zeta transform.] 
  For each $s=0,\ldots,n$, set $g_0^{(s)}(X)=[X\in\scr F][|X|=s]$. 
  For each $j=1,\ldots, n$ and $s=0,\ldots, n$, set
    \[ 
       g_j^{(s)}(X) = [j\in X]      g_{j-1}^{(s)} (X\setminus\{j\}) + 
                                    g_{j-1}^{(s)} (X)\,.
    \]
  [At this point it holds that $g_n^{(s)}(X)=(f\zeta^{(s)})(X)$ 
   for all $0\leq s\leq n$.]
\item[{\bf D4:}] [Evaluate zeta transform of $d^{(s)}$.] 
  For each $s=0,\ldots, n$, set $q(1,s)=g_n^{(s)}(X)$. 
  For each $i=2,\ldots,k$ and $s=0,\ldots,n$, set 
  $q(i,s)= \sum_{t=0}^s q(i-1,s-t) g_n^{(t)}(X).$
\item[{\bf D5:}] [Ranked Moebius transform.] 
  For each $s=0,\ldots,n$, set $h_0^{(s)}(X)=q(k,s)$. 
  For each $j=1,\ldots,n$ and $s=0,\ldots,n$, set 
    \[ 
       h_j^{(s)}(X)= 
            -[j\in X] h_{j-1}^{(s)} (X\setminus\{j\}) + h_{j-1}^{(s)} (X)\,.
    \]
  [At this point it holds that $h_n^{(s)}(X)=d^{(s)}(X)$ 
   for all $0\leq s\leq n$.]
\item[{\bf D6:}] Output $X$ and $h_n^{(|X|)}(X)$.
\item[{\bf D7:}] For each $j\notin X$, insert $X\cup \{j\}$ into $\scr L(r+1)$.
\item[{\bf D8:}] If $\scr L(r)$ is empty, then increment $r\leq n$ until $\scr L(r)$
  is nonempty; terminate if $r=n$ and $\scr L(n)$ is empty.
\item[{\bf D9:}] Go to D2.
\end{itemize}}

\subsection{Packings}
According to \eqref{eq: p=dzeta}, to compute $p(X)$ it suffices to 
zeta-transform the partition number. This amounts to running
Algorithm~Z after Algorithm~D. (For a different approach, see \cite{BHK08}.)

\section{Applications}

\vskip-0.3cm
\subsection{The number of dominating sets in sparse graphs}
\label{sec: comb}
This section is purely combinatorial. Let $\scr D$ denote the
dominating sets of a graph. A complete graph has $2^n-1$ dominating
sets, and sparse graphs can have almost as many: the $n$-star graph
has $2^{n-1}$ dominating sets and average degree less than $2$. Thus
we ask how large $|\scr D|$ can be for graphs with bounded maximum
degree. An easy example is provided by the disjoint union of complete
graphs of order
$\Delta+1$: every vertex subset that includes at least
one vertex from each component is dominating, so $|\scr D| =
(2^{\Delta+1}-1)^{n/(\Delta+1)}$. We shall show that this is in fact
the largest possible $\scr D$ for graphs of maximum degree
$\Delta$. Our analysis is based on the following intersection theorem.

\begin{lem}[Chung \emph{et al.} \cite{CFGS86}]
\label{lem: chung}
Let $U$ be a finite set with subsets $P_1,\ldots, P_m$ such that every
$u\in U$ is contained in at least $\delta$ subsets. Let\/ $\scr F$ be a
family of subsets of\/ $U$. For each\/ $1\leq \ell \leq m$, define the
projections $\scr F_\ell= \{\, F\cap P_\ell \colon F\in \scr F\,\}$. Then 
\[ 
 |\scr F|^\delta \leq \prod_{\ell=1}^m |\scr F_\ell|\,.
\]
\end{lem}

\begin{thm}\label{thm: dom}
  The number of dominating sets of an $n$-vertex graph with maximum
  degree $\Delta$ is at most\/ \( (2^{\Delta+1}-1)^{n/(\Delta+1)}.\)
\end{thm}

\begin{pf}
  Let $G=(V,E)$ be a graph with $|V|=n$ and maximum degree $\Delta$.
  For each $v\in V$, let $A_v$ be the closed neighbourhood around
  vertex $v$, 
  \begin{equation}\label{eq: Ai def} A_v = \{v\}\cup
    \{\,u\in V\colon uv\in E\,\}\,.
  \end{equation}
  Next, for each $u\in V$ with degree $d(u)<\Delta$, add $u$ to
  $\Delta-d(u)$ of the sets $A_v$ not already containing $u$ 
  (it does not matter which). Let $a_v=|A_v|$ 
  and note that $\sum_v a_v=(\Delta+1) n$.
  
  We want to apply Lemma~\ref{lem: chung}. To this end, let $U=V$ and $m=n$.
  By construction, every $u\in V$ belongs to exactly $\delta=\Delta+1$ 
  subsets $A_v$. To get a nontrivial bound on $\scr D$ we need to bound 
  the size of $\scr D_v = \{\, D\cap A_v \colon D\in\scr D\,\}$. 
  Every $D\cap A_v$ is one of the $2^{a_v}$ subsets of $A_v$, but none 
  of the $D\cap A_v$ can be the empty set, because either $v$ or one of 
  its neighbours must belong to the dominating set $D$. Thus 
  $|\scr D_v|\leq 2^{a_v}-1$. By Lemma~\ref{lem: chung}, we have
  \begin{equation}\label{eq: from chung lemma}
     |\scr D|^{\Delta+1} \leq \prod_v (2^{a_v}-1)\,.
  \end{equation}
  Since $x\mapsto \log\,(2^x-1)$ is concave, Jensen's inequality gives
  \[
  \frac{1}{n}\sum_v \log\,(2^{a_v}-1) \leq 
  \log\,(2^{\sum_v a_v/n}-1) = \log\,(2^{\Delta+1}-1)\,.
  \] 
  Taking exponentials and combining with \eqref{eq: from chung lemma} 
  gives $|\scr D|^{\Delta+1} \leq (2^{\Delta+1}-1)^n$.
\end{pf}

\vskip-0.3cm
\subsection{Domatic Number}

We first observe that a graph can be packed with $k$ dominating sets
if and only if it can be packed with $k$ \emph{minimal} dominating
sets, so we can consider $k$-packings from $\min \scr D$ instead of
$\scr D$. This has the advantage that $\min\scr D$ can be listed
faster than $2^n$.

\begin{lem}[Fomin \emph{et al.} \cite{FGPS05}]
  Any $n$-vertex graph has at most\/ $O^*(1.7170^n)$ minimal dominating
  sets, and they can be listed within that time bound.
\end{lem}

\begin{thm}\label{thm: applications}
  For an $n$-vertex graph $G$ with maximum degree $\Delta$ we can
  decide in time 
  \[
  O^*\bigl((2^{\Delta+1}-2)^{n/(\Delta+1)}\bigr) 
  \]
  whether $G$ admits a packing with $k$ dominating sets.
\end{thm}
\begin{pf}
  We use Algorithm~D with $\scr F= \min \scr D$. By the above lemma,
  we can complete Step~D1 in time $O^*(1.7170^n)$. The rest of the
  algorithm requires time $O^*(|\upset \min\scr D|)$. Since
  every superset of a dominating set is itself dominating,
  $\upset\min\scr D$ is a sub-family of $\scr D$ (in fact, it is
  exactly $\scr D$), so Theorem~\ref{thm: dom} bounds the total running
  time by 
  \[
  O^*\bigl((2^{\Delta+1}-1)^{n/(\Delta+1)}\bigr)\,.
  \]
  
  We can do slightly better if we modify Algorithm~D in Step~D7 
  to insert $X\cup \{j\}$ only if it excludes at least one vertex for each
  closed neighbourhood.  Put otherwise, we insert $X\cup \{j\}$
  only if the set $V\setminus (X\cup \{j\})$ dominates the graph $G$.  
  The graph then has Domatic Number at least $k+1$ if and only if the
  algorithm reports some $X$ for which $d(X)$ is nonzero.  The
  running time can again be bounded as in Theorem~\ref{thm: dom} but now
  $D\cap A_v$ can neither be the empty set, nor be equal to
  $A_v$. Thus the application of Lemma~\ref{lem: chung} can be
  strengthened to yield the claimed result.
\end{pf}

\vskip-0.3cm
\subsection{Chromatic Number}

Our first argument for Chromatic Number is similar; we give a stronger
and slightly more complicated argument in \S\ref{sec: chromatic
  via bip}.

We consider the independent sets $\scr I$ of a graph.  An independent
set is not necessarily dominating, but it is easy to see that a
\emph{maximal} independent set is dominating. Moreover, the
Moon--Moser bound tells us they are few, and Tsukiyama \emph{et al.} tell us how to list them with polynomial delay:

\begin{lem}[Moon and Moser \cite{MM65}; Tsukiyama \emph{et al.} \cite{TIAS77}]
  Any $n$-vertex graph has at most\/ $O^*(1.4423^n)$ maximal independent
  sets, and they can be listed within that bound.
\end{lem}

\begin{thm}\label{thm: k independent sets}
  For an $n$-vertex graph $G$ with maximum degree $\Delta$ we can
  decide in time 
  \[
  O^*\bigl((2^{\Delta+1}-1)^{n/(\Delta+1)}\bigr) 
  \] 
  whether $G$ admits a covering with $k$ independent sets.
\end{thm}

\begin{pf}
  It is easy to see that $G$ can be covered with $k$ independent sets
  if and only if it can be covered with $k$ \emph{maximal} independent
  sets, so we will use Algorithm~C on $\max\scr I$. Step~C1 is
  completed in time $O^*(1.4423^n)$, and the rest of the algorithm
  considers only the points in $\upset \max \scr I$, which all belong
  to $\scr D$. Again, Theorem~\ref{thm: dom} bounds the total running
  time.
\end{pf}

\vskip-0.3cm
\subsection{Chromatic Number via bipartite subgraphs}
\label{sec: chromatic via bip}

We can do somewhat better by considering the family $\scr B$ of 
vertex sets of induced \emph{bipartite} subgraphs, that is, the family 
of sets $B\subseteq V$ for which the induced subgraph $G[B]$ is bipartite. 
As before, the literature
provides us with a nontrivial listing algorithm:

\begin{lem}[Byskov and Eppstein \cite{Byskov-thesis}]
  Any $n$-vertex graph has at most\/ $O^*(1.7724^n)$ maximal induced
  bipartite subgraphs, and they can be listed within that bound.
\end{lem}

The family $\max\scr B$ is more than just dominating, which allows us
to use Lemma~\ref{lem: chung} in a stronger way.

\begin{thm}
\label{thm: c}
  For an $n$-vertex graph of maximum degree $\Delta$ it holds that
  \[
  |\upset\max\scr B| \leq (2^{\Delta+1}-\Delta-1)^{n/(\Delta+1)}\,.
  \]
\end{thm}
\begin{pf}
  Let $G=(V,E)$ be a graph with $|V|=n$ and maximum degree $\Delta$.
  Let $\scr F=\upset\max\scr B$. Let $A_v$ be as in
  \eqref{eq: Ai def}. With the objective of applying
  Lemma~\ref{lem: chung}, we need to bound
  the number of sets in $\scr F_v=\{\,F\cap A_v \colon F\in \scr F\,\}$. 

  Assume first that $G$ is $\Delta$-regular.  Let
  $A_v=\{ v, u_1,\ldots, u_\Delta \}$. We will rule out $\Delta+1$
  candidates for $F\cap A_v$, namely
   \begin{equation}
    \label{eq: bip ruled out} \emptyset, \{u_1\},
    \ldots, \{u_\Delta\}\notin \scr F_v\,.
    \end{equation} 
  This then shows
  that $|\scr F_v|\leq 2^{\Delta+1}-\Delta-1$ and thus the bound follows
  from Lemma~\ref{lem: chung}.

  To see that \eqref{eq: bip ruled out} holds, observe that $F\in\scr F$ 
  contains a $B\subseteq F$ such that the induced subgraph $G[B]$ is maximal
  bipartite. To reach a contradiction, assume that there exists 
  a $v\in V$ with $F\cap A_v\subseteq \{u_\ell\}$. Since $B\subseteq F$, 
  we have $B\cap A_v\subseteq \{u_\ell\}$, implying that $v$ does not 
  belong to $B$, and that at most one of its neighbours does. 
  Consequently, $G[B\cup\{v\}]$ is also bipartite, and $v$ belongs to a
  partite set opposite to any of its neighbours. This contradicts
  the fact that $G[B]$ is maximal bipartite.

  To establish the non-regular case, we can proceed as in the proof
  of Theorem~\ref{thm: dom}, 
  adding each $u\in V$ with $d(u)<\Delta$ to some $\Delta-d(u)$
  of the sets $A_v$ not already including $u$.
  Note that by adding $y$ new vertices to $A_v$ originally containing $x$ vertices,
  we get $|\{F\cap A_v:F\in \scr F\}|\leq 2^y(2^x-x-1)$. Next, since 
  $2^y(2^x-x-1)\leq 2^{y+x}-(y+x)-1$ for all non-negative integers $y,x$ and
  $\log\,(2^x-x-1)$ is a concave function, the bound follows as before via Jensen's inequality.  
\end{pf}

\begin{thm}\label{thm: k independent sets with bip}
  For an $n$-vertex graph $G$ with maximum degree $\Delta$ we can
  decide in time 
  \[
  O\bigl((2^{\Delta+1}-\Delta-1)^{n/(\Delta+1)}\bigr) 
  \] 
  whether $G$ admits a covering with $k$ independent sets.
\end{thm}
\begin{pf}
  When $k$ is even, 
  it is easy to see that $G$ can be covered by $k$ independent
  sets if and only if it can be covered by $k'=k/2$ maximal bipartite
  sets, so we will use Algorithm~C on $\max\scr B$ and investigate
  whether $c(V)\neq 0$.
  
  When $k$ is odd, we again use Algorithm~C with $k'=(k-1)/2$
  maximal bipartite sets, but this time we check whether an $X$ 
  is output such that both $c(X)\neq 0$ and $V\setminus X$ is 
  independent in $G$.
 
  In both cases the running time bound follows from Theorem~\ref{thm: c}.
\end{pf}

\section{Concluding Remarks}
\label{sec: con}

Since the presented improvements on running time bounds are 
modest, one can ask whether this is because of weak bounds or 
because of inherent limitations of the technique. We observe that 
the running time bounds in Theorems \ref{thm: applications}, 
\ref{thm: k independent sets}, and \ref{thm: k independent sets with bip} 
are met by a disjoint union of complete graphs of order $\Delta+1$.
Thus, either further trimming or splitting into connected components
is required for improved algorithms in this context. 

We chose to demonstrate the technique for Chromatic and Domatic Number
since these are well-known and well-studied. To briefly demonstrate 
some further application potential, more artificial problem 
variants such as determining if a $\Delta$-regular graph has domatic
number at least $\Delta/2$, or if the square of a
$\Delta$-regular graph has chromatic number at most
$3\Delta/2$, admit stronger bounds. For example, 
if $G$ has domatic number at least $d$, $d$ even, then its vertices 
can be partitioned into two sets, both of which contain $d/2$ dominating 
sets. This suggests the following meet-in-the-middle strategy. 
Run Algorithm~D with $\scr F$ equal to all dominating sets and $k=d/2$, 
but modify Step~D7 to insert $X\cup \{j\}$ only 
if $|A_v\setminus (X\cup\{j\})|\geq d/2$ holds
for all vertices $v$. At termination, we check
whether the algorithm has output two sets $X$ and $Y$ such
that $X\cup Y=V$ and $d(X),d(Y)>0$. (For example, one can
check for duplicates in a table with entry $\{X,V\setminus X\}$ 
for each output $X$ with $d(X)>0$.)
This algorithm variant considers only sets with many 
forbidden intersections with the neighbourhoods of vertices, 
which translates into stronger bounds via Lemma~\ref{lem: chung}.


\end{document}